\documentclass[final]{aipproc}

\usepackage{amssymb}
\usepackage{epsfig,rotating}
\layoutstyle{6x9}

\def\pom{\mathrm{I\!P}}

\begin{document}

\title{Double Pomeron Exchange: \\ from the ISR to the LHC}


\author{Michael Albrow}{
  address={Fermi National Accelerator Laboratory\
  P.O.Box 500, Wilson Road, Batavia, IL 60510, USA}
\\
}

\begin{abstract}
 I discuss Double Pomeron Exchange processes from their first observation at the CERN Intersecting Storage Rings,
 focusing on glueball searches, through the observations of exclusive $\chi_{c0}, \gamma\gamma$ and di-jets at the
 Tevatron, to prospects at the LHC for exclusive Higgs boson production.
\end{abstract}

\maketitle


When two protons collide at high energies they may emerge with small loss of energy (in the center of mass frame, at most a few percent), 
having created a low mass system $X$ of particles at central rapidity. There is then a rapidity gap of at least 
$\Delta y = 3$ (preferably $> 4$) units between $X$ and each proton $p$. The $t$-channel exchanges between $p$ and $X$ can 
only be color singlets with spin $J$, or effective spin $\alpha(t), \geq$ 1; in the Standard Model (SM) 
only the photon, $\gamma$, or the pomeron, $\pom$. The pomeron is at leading order a pair of gluons [$gg$] in a CP = ++ state. 
A third possibility is the odderon, at least three gluons with C = -1; there is no good evidence for this yet,
and I shall ignore it. 

Two-photon exchange has been seen at the Tevatron (CDF)~\cite{cdfee,mumu34,cdfz} as $\gamma\gamma \rightarrow e^+e^-$ 
and $\mu^+ \mu^-$; it is a QED process with small corrections from the proton form factor. Photon-pomeron fusion, 
or photoproduction, was a major topic at the HERA $e-p$ collider, recently also observed by CDF~\cite{mumu34} in the 
$J/\psi$ and $\psi(2S)$ states. CMS has now shown candidates for $J/\psi$ and $\Upsilon$ photoproduction. The LHC 
opens up the possibility of measuring $Z$-photoproduction; in the SM it is marginally too small ($\lesssim$ 50 fb), but could 
be enhanced by additional BSM loops that couple to $\gamma, Z$ and $\pom$. CDF published~\cite{cdfz} an upper limit a 
factor $\sim$ 1000 above the SM prediction.

    I now restrict myself to $\pom \pom$ interactions, or double pomeron exchange; see ~\cite{acf} for a recent review. 
    This is $p+p \rightarrow p+X+p$ with \emph{both} protons having Feynman $x_F \gtrsim$ 0.95, and with rapidity gaps $\Delta
    y \gtrsim 3$ between the protons and $X$. These are strongly correlated features and one can require either or (better)
    both, but there is no sharp \emph{separation} between pomeron and other exchanges.
    They have a special feature in that they do not involve 
    valence quarks (any quarks present must have evolved in as virtual pairs via $g \rightarrow q \bar{q}$ ), and the properties of $X$ should 
    be independent of the colliding hadrons, no matter whether they are $pp$, $\Omega^-  \bar{\Omega}^+$ or $\pi^+\pi^+$. 
    That is the supposition, not however tested by experiment.  
    (All fixed target experiments have too low $\sqrt{s}$ to study $\pom \pom \rightarrow X$ 
    with little background). Furthermore, in $\pom + \pom \rightarrow X$, the state $X$ has tightly constrained 
    properties; it must have CP = ++, Q = 0, no net flavor, and spin $J$ even (mostly $J$ = 0). Together 
    with its glue dominance this makes D$\pom$E an excellent channel for spectroscopy when M($X) \lesssim$ few GeV, 
    and for studying QCD phenomena (and the pomeron itself) at large M($X$). 
     Central masses M($X$) extend up to $\sim$ 4\% of $\sqrt{s}$ (corresponding to $x_F = 0.96$), and at the LHC reach the electroweak
     sector. 
     The Higgs boson has the quantum numbers of the vacuum, obeys all the rules for  D$\pom$E, and can be produced "exclusively" 
     (meaning no other particles are produced) as $p+p \rightarrow p + H + p$. Just as pomeron exchange in $p+p \rightarrow p + X$
      is called "diffractive excitation of a proton", D$\pom$E can be called "diffractive excitation of the vacuum". Any states with 
      allowed quantum numbers are present virtually in the vacuum and can be "kicked into reality" in the collision 
      of two protons. While in $\gamma\gamma$ processes the impact parameter in the collision is usually several fm, with the
      scattered protons at correspondingly small $|t|$, the most likely impact parameter, $b$, in $p+p \rightarrow p + H + p$ is
      intermediate: $b \lesssim 1$ fm but not $b \sim 0$, because in that case of maximum overlap the rapidity gap survival
      probability is minimal. 
       The Higgs boson couples to the protons only weakly through (mainly) the loop $gg \rightarrow t\bar{t} \rightarrow H$, 
      which is the dominant inclusive Higgs production mechanism. Normally the protons are left colored after the 
      gluon annihilation and a large number of particles are produced. However it is possible for another gluon of 
      complementary color to be exchanged, for the protons to remain in their ground state, and even to have no hadrons 
      created: $p + p \rightarrow p + H + p$, or $\pom + \pom \rightarrow H$.
One pays a big price, $\sim 10^{-3} - 10^{-4}$ in cross section, for these conditions, but if even a few events are observed, 
with the protons well measured and the $H$-decay detected, the payoff will be big. For the SM Higgs, whether it decays 
to $b\bar{b}$, $\tau^+\tau^-$, $W^+W^-$ or even $ZZ \rightarrow \mu^+\mu^-\nu\bar{\nu}$, the central 
mass M($X$) = M($H$) can be measured with $\sigma(M) \sim$ 2 GeV/c$^2$ from the missing mass to the two protons ~\cite{albrow}. The 
background should be small~\cite{fp420}, and the CP must be ++, the spin $J=$0 or 2 (and these can be distinguished 
from the proton azimuthal angles~\cite{kkmr}), and the state width can be measured (if $\Gamma \gtrsim 3$ GeV). If there are multiple Higgses, 
as in SUSY models, they could be separated even if only a few GeV apart. However the expected cross section is 
low, $\sim$ 1 - 10 fb for the SMH in the Durham model~\cite{khoze}, and uncertain (it is higher in the MSSM). Fortunately the calculations can be checked 
with other D$\pom$E reactions that can be measured at the Tevatron, but before describing these, I return to the 
origins of D$\pom$E physics at the CERN Intersecting Storage Rings, ISR.

    The ISR had first collisions in January 1971, and gave us a large step in $\sqrt{s}$ from 7.6 GeV to 63 GeV. Single 
    diffractive excitation, previously only of low mass states (resonances),m was found to extend up to about 
    13 GeV/c$^2$ (for $x_F >$ 0.96), well above 
    the resonance region. This was well described by Regge theory (our best approach to strong reactions pre-QCD) 
    in terms of "triple Regge diagrams". A pomeron "emitted" (not really!) from a proton interacts with the other 
    proton: $\pom + p \rightarrow X$, and the $\pom p$ total cross section is related by the optical theorem to ``$\pom +
    p$'' 
    elastic scattering with another $t$-channel exchange. For low M($X$) the latter is a reggeon (virtual $f_2$ exchange), 
    but at high M($X$) it should be another pomeron, the diagram including a triple pomeron coupling $g_{\pom \pom \pom}$. 
    A higher order diagram (double-triple pomeron or quintuple pomeron) is then implied, corresponding to the D$\pom$E process $\pom + \pom \rightarrow X$. 
    Several experiments at the ISR sought D$\pom$E, but the evidence was slow to accumulate.
The total rapidity  range is 2.ln$(2E_{beam}/m_p) =  6.4 (8.4)$ at $\sqrt{s}$ = 23 (63), so the possibility of having two 
gaps $\Delta y \gtrsim 3$ is limited to the higher energies; one also needed detectors in both the central and very 
forward regions. Eventually D$\pom$E became established, for two (unidentified, but presumed to be $\pi^+\pi^-$) hadrons, 
with a cross section as expected from Regge phenomenology tuned to single diffraction, elastic and total cross 
sections. If one fixes the forward rapidity gaps to $\Delta y$ = 3 (say), the cross section rises with $\sqrt{s}$ 
as the allowed central region expands. If on the other hand one fixes the central region for the pions to be $|y| < 1.0$ or $< 1.5$ 
the cross section decreases slowly as the gaps get longer. The Split Field Magnet experiments eventually observed resonant $f_0/f_2$ 
signals in the $\pi^+\pi^-$ (assumed) spectrum.
    In the last days of the ISR forward drift chambers were added to the Axial Field Spectrometer~\cite{afs} and provided measurements of $\pom \pom \rightarrow \pi^+\pi^-$ with high statistics, 
    as well as $K^+K^-, p\bar{p}$, and $4\pi$ with identified particles. 
    
    Consider the known particles with the 
    allowed $X$-quantum numbers. They are the very broad $f_0(600)$ (or $\sigma), f_0(980), f_2(1270), f_0(1400) ...  \chi_{c0},$ and $ \chi_{b0}$. 
    Not yet known, but with the allowed quantum numbers, are the Higgs boson(s) and perhaps graviton (in theories with 
    extra dimensions, a spin 2 massive graviton $G$ could exist and be produced in D$\pom$E). The AFS $\pi^+\pi^-$ 
    spectrum showed the $f_0(980)$ as a dip, above a very broad scalar ($J^P = 0^+$) distribution likely to be 
    dominated by the $\sigma$, and a small $f_2(1270)$ under a broad scalar, probably $f_2(1400)$. Interestingly 
  the AFS also saw~\cite{afs}  $\alpha\alpha \rightarrow \alpha + \pi^+\pi^- + \alpha$ with the same $\pi^+\pi^-$ spectrum, 100\% 
    background-free as the $\alpha$'s must have been coherently scattered.
    
   The ISR gave way to the Sp$\mathrm{\bar{p}}$S collider, and the focus on $W,Z$ and jet physics did not leave much room 
   for D$\pom$E studies. Proton trackers were added~\cite{ua8} to the UA2 central detector and D$\pom$E was observed, but with poor 
   mass resolution and small statistics. Forward gap triggers allowed a study~\cite{ua1} with the UA1 detector of higher M($X$)
    events, showing some soft jettiness and evidence that the charged multiplicity is higher (and rises faster with
    M($X$)) than in $e^+e^-$ 
    collisions, which is not surprising.

The next step came with CDF (the Collider Detector at Fermilab) at the Tevatron with $\sqrt{s}$ = 1960 GeV, allowing 
central masses up to about 80 (100) GeV 
with $x_F > 0.96 (0.95)$.  So far there have been no results on low mass spectroscopy, because 
D$\pom$E triggers were not in place until recently. Neither CDF nor DZero now have forward proton spectrometers, but recently we 
developed a trigger in CDF based on forward rapidity gaps, vetoing on particles with 2.1 $< |\eta| <$ 5.7 on each side. Either the protons 
went down the beam pipe, or they dissociated into a low mass (M $\lesssim$ 2 GeV) state; 
in either case the events are $\pom \pom \rightarrow X$ (with small contributions from $\gamma \pom$ and $\gamma\gamma$
exchanges). The most efficient luminosity for this data is when the average number of inelastic collisions per bunch crossing is $\sim 1$
i.e. $L \sim 4 \times 10 ^{31}$ cm$^{-2}$s$^{-1}$ (these days the beams are usually dumped at higher $L$), and as the cross
sections are several $\mu$b the event rate is high enough to collect millions of events in a few hours. In 
the low mass region exclusive hadron pairs ($\pi^+\pi^-, K^+K^-, \phi\phi, ...$) etc. are interesting for spectroscopy, and it may be
possible to see hadronic decays of exclusive $\chi_{c}$ states, which would be important for testing calculations of 
exclusive $H$ production. At high masses, M($X) \sim$ 50 GeV, many studies of event shapes, jets,
multiplicities, Drell-Yan pairs, Bose-Einstein correlations, and so on can teach us about pomeron interactions, provide data
to test predictions (e.g. by \textsc{phojet}) and learn about backgrounds to exclusive Higgs. 

Exclusive dijets, $p + \bar{p} \rightarrow p + JJ + \bar{p}$ were studied in CDF~\cite{cdfjj} by triggering on two jets with a high-$x_F$ antiproton and
requiring a rapidity gap on the $p$-side (there was no $p-$detector). Off-line one found events with most of the central energy
contained in two jets, in excess of expectations unless the exclusive process $\pom +\pom \rightarrow J+J$ is included. The
\textsc{exhume} program, based on the ``Durham process", calculates this as $gg \rightarrow J+J$ with another gluon exchange
to cancel the color, as in exclusive Higgs production, and the data are in fair agreement (a factor of ``a few" in theory and
experiment). A model with the pomeron as an object with a multi-$q/\bar{q}/g$ structure such as \textsc{pomwig} does not give
exclusive dijets and fails to reproduce the data. The CDF exclusive dijet study extends up to M($JJ) \sim$ 120 GeV in only
300 pb$^{-1}$ of luminosity. Recently DZero has seen 26 exclusive dijet event candidates~\cite{dojj} with M($JJ) >$ 100 GeV 
(30 pb$^{-1}$ of data). 

While the exclusive dijet cross section measurement has model dependence, and there is no \emph{absolute} distinction between 
exclusive and inclusive
dijets, exclusive $\chi_c$ production is well-defined; it is $p + \bar{p} \rightarrow p + \chi_c + \bar{p}$ with \emph{no other
hadrons} in the final state. The diagram is the same as exclusive $H$ production with the top-loop replaced with a charm-loop,
and the $\chi_{c0}$ has the same quantum numbers, which makes it a clean comparison. It was observed by CDF~\cite{mumu34} with the $\chi_c$
central, with $d\sigma/dy|_{y=0} = 76\pm10(\mathrm{stat})\pm 10(\mathrm{syst})$ nb in $\chi_c \rightarrow J/\psi+\gamma$ with
$J/\psi \rightarrow \mu^+\mu^-$, in agreement with the Durham prediction~\cite{khoze}. Unfortunately the photon has low
energy ($\sim$ 200 MeV) so the M($J/\psi+\gamma$) resolution does not distinguish $\chi_c$ states; the $\chi_{c0}(3415)$ state
should dominate but the $\chi_{c1}(3510)$ and $\chi_{c2}(3556)$ states have 30(17)$\times$ higher branching fractions respectively. It should be
possible to distinguish these states using hadronic decays, which have much better mass resolution. About 6\% of $\chi_{c0}$ decays
are to $h^+h^-$ or $2(h^+h^-)$, where $h=\pi$ or $K$. The width $\Gamma(\chi_{c0})$ = 10 MeV and the mass resolution in CDF 
is similar; the main issue is continuum background. The CDF observation of $\chi_c \rightarrow J/\psi + \gamma$ does not suffer from either combinatorial or continuum backgrounds; the only
significant ``background" to $p+\chi_c + \bar{p}$ is undetected dissociation e.g. $p+\chi_c + \bar{p}\pi^+\pi^-$,
with the dissociation products not detected in the forward Beam Shower Counters, BSC.
We will also search for exclusive open charm: $D^+D^-, D^0 \bar{D}^0, D_s^0 \bar{D}_s^0$ in hadronic modes.

CDF also published a search for $\pom \pom \rightarrow \gamma\gamma$~\cite{exclgg}, with $E_T(\gamma) >$ 5 GeV, which proceeds through an 
intermediate quark loop, and is the closest control process to exclusive Higgs as the final state does not have
strong interactions. Unfortunately the cross section is small; Ref.~\cite{durhgg} predicted $36^{+72}_{-24}$ fb corresponding
to 0.8$^{+1.6}_{-0.5}$ events, and three candidates were found. The $E_T$ threshold has since been decreased to $\sim 2.5$ GeV
and there are many more candidates, but the $\pom +\pom \rightarrow \pi^0\pi^0$ background has to be understood; it is an
interesting  channel in itself. 

RHIC has now entered the field of D$\pom$E with forward proton measurements in STAR, reported at this meeting by
Guryn~\cite{rhic}.

At the LHC D$\pom$E studies will be very interesting, both in the low mass exclusive regime and at 
high masses M($X) \gtrsim$ 200 GeV (quite apart from the $H$ search). At this meeting 
Schicker showed some preliminary ALICE data~\cite{alice}. However at present all experiments are handicapped by having very poor forward coverage.
There are $\theta = 0^\circ$ calorimeters (ZDC) for $\gamma$ and neutrons ($K^0_L/ n \lesssim 10^{-2}$), and in CMS the HF 
calorimeters have
$|\eta| < 5.2$, with CASTOR on one side having $5.2 < \eta < 6.4$. About 3 units of $\eta$ on both sides 
are uninstrumented,
exactly where most rapidity gaps are in diffractive interactions. Events with hadrons between two clean 
gaps $\Delta y \gtrsim 4$ are
confined to $-2.4 < y < +1.2$ (not minding whether the $p$ dissociated). These gaps can be simply covered 
with sets of scintillation counters (FSC = Forward Shower Counters)~\cite{fsc} along the beam pipes 
out to $|z| \sim 140$ m. ALICE is installing counters and they are proposed for CMS. Among many other things 
(including a $\sigma_{\mathrm{inel}}$ measurement and single diffraction) they allow a D$\pom$E
trigger based on vetos in FSC, ZDC, CASTOR and HF, together with some minimal central activity. These 
select $p+X+p$ events,
without however detecting the protons, which can be done with the TOTEM detectors in some conditions 
(not yet done). They also
allow D$\pom$E studies with low mass proton dissociation ($p^*+X+p^*$), by triggering on hits in the FSC (both sides) with a veto on HF and a
small central energy deposit or track. While this should give a clean  sample, more information ($t_1,t_2,\Delta\phi$) would come from detecting the
scattered protons, as could be done in CMS+TOTEM. During the expected special high-$\beta^*$ (90 m) run for TOTEM, the luminosity will be low
and perhaps the best use of the time for CMS would be D$\pom$E with a combined trigger (on each side a $p$ 
and $\overline{ZDC.FSC.HF}$ 
with some central activity). 

Comparing D$\pom$E in baryon-baryon and meson-meson collisions at the LHC \emph{might} be possible, with $n$ in each ZDC, and FSC hits
from ``$\pi^+$'' with adjacent gaps. Better, study $J/\psi + J/\psi \rightarrow J/\psi +X+ J/\psi$ 
collisions at future $e^+e^-, e^-e^-$(better!) or $\mu^+\mu^-$ colliders to test whether pomerons from 3$q$ and $q\bar{q}$ hadrons are different.

  To conclude, D$\pom$E is a special kind of interaction: strong and almost pure glue, with constrained quantum 
  numbers and ``clean'', and few
  allowed \emph{exclusive} central states: $f_{0,2}, \chi_{c,b}, \gamma\gamma, JJ, H$. The low mass end addresses glueball and meson
  spectroscopy, and we have hardly scratched the surface. The high mass end addresses pomeron structure, jets, and much else. We could
  get much more data at the Tevatron, and we are just starting at the LHC, where we should push for maximal forward coverage (FSC) and forward gaps in level-1 triggers.
  
  I thank the DOE for support, and especially Valery Khoze, Alan Martin and Misha Ryskin for discussions.



\bibliographystyle{aipproc}   



\end{document}